\documentclass[preprint,12pt]{elsarticle}
\usepackage{amssymb,epsfig,graphicx,multirow,bm,dcolumn}
\journal{JMMM}
\begin{document}
\begin{frontmatter}

\title{Modulations in magnetostructural coupling in C and Sn deficient Mn$_3$SnC}
\author[gu]{V. N. Gaonkar}
\author[tifr]{E. T. Dias}
\author[barc]{A. Das}
\author[tifr]{A. K. Nigam}
\author[gu]{K. R. Priolkar \corref{krp}}\ead{krp@unigoa.ac.in}
\cortext[krp]{Corresponding author}
\address[gu]{Department of Physics, Goa University, Taleigao Plateau, Goa 403206 India}
\address[tifr]{Tata Institute of Fundamental Research, Dr. Homi Bhabha Road, Colaba, Mumbai 400005 India}
\address[barc]{Solid State Physics Division, Bhabha Atomic Research Centre, Trombay, Mumbai 400085 India}

\begin{abstract}
The structural and magnetic interactions in stoichiometric (Mn$_3$SnC), carbon deficient (Mn$_3$SnC$_{0.8}$) and tin deficient (Mn$_3$Sn$_{0.85}$C) antiperovskite compounds are studied using x-ray absorption fine structure spectroscopy and neutron diffraction. The study confirms the presence of local structural distortions only around Mn atoms in the antiperovskite compounds irrespective of their stoichiometry. The distortions in the Mn$_6$C octahedra are such that only Mn atoms are displaced from their crystallographic positions resulting in long and short Mn--Mn bonds. These long and short Mn--Mn bonds are responsible for presence of ferromagnetic and antiferromagnetic moments on Mn atoms. The C deficiency at the center of the octahedra increases the strain on the Mn$_6$C octahedra and results in a wide variation of Mn--Mn bond distances as a function of temperature and large hysteresis in magnetic properties. \emph{On the other hand, Sn deficiency tends to relax strain by giving more space for the octahedra to distort leading to temperature independent Mn--Mn bond distances}.
\end{abstract}
\date{\today}

\end{frontmatter}

\section{Introduction}
Mn$_3$SnC belongs to a class of compounds called antiperovskites which display a slew of interesting properties like giant negative thermal expansion \cite{Hamada2011109,Takenaka2005}, near--zero temperature coefficient of resistance \cite{takenaka201198,Lei201199}, large magnetocaloric effect (MCE) \cite{Tohei200394,Lewis200393,Aczel201490}, giant magnetoresistance (GMR) \cite{Kamishima200063,Li200572,Zhang2014115}, etc.  These compounds crystallize in the famous cubic perovskite structure (ABX$_3$) but with Mn atoms occupying the centers of cube faces (X sites) and forming a octahedral cage around a non metal atom  like C or N which occupy the B sites \emph{and have attracted attention over past several years \cite{fruchart197844,kaneko198756,takenaka201415}}.

Mn$_3$SnC itself exhibits a large magnetic entropy change near room temperature  ($\Delta S_{max} \sim 10.37$ J/Kg--K under a magnetic field change of \emph{2T}) which is associated with a first order phase transition from a low volume cubic paramagnetic high temperature phase to a high volume cubic low temperature phase with complex magnetic order \cite{Wang200985,Dias201548}. Here in, two of the three manganese atoms align antiferromagnetically with their spins pointing in the $x - y$ plane while the third Mn atom carries a pure ferromagnetic moment of 0.7$ \pm $0.4$ \mu_{B} $ along the 001 direction \cite{Dias201548}. Therefore, unlike the cubic crystal structure, the tetragonal antiferromagnetic unit cell ($a\sqrt{2}$, $a\sqrt{2}$, $a$) of Mn$_3$SnC is generated using a propagation vector $k = \left [{1\over 2}, {1\over 2}, 0 \right ]$. Recent time dependent magnetization studies on Mn$_3$SnC have shown that upon transformation, all the three Mn spins first order ferromagnetically along 001 direction and only later, two of them flip to order antiferromagnetically in the $x - y$ plane \cite{Cakir201796}. Such a flipping of Mn spins causes mechanical strain which is in addition to the strain produced due to change in unit cell volume at the phase transition.

The strain on the Mn$_6$C functional units can be effectively modulated by changing the A site atom in such Mn based antiperovskites and even render the resulting compounds functional. In Mn$_3$Cu$_{1-x}$Ge$_x$N critical substitution of Cu by Ge results in large negative thermal expansion at about room temperature \cite{Sun200942}. On the other hand, replacement of Sn by Mn in Mn$_{3+x}$Sn$_{1-x}$C results in a change in nature of transition from first order to second order \cite{Wang2012}. The A site induced strain also modifies the magnetic and magnetocaloric properties of Mn$_3$Ga$_{1-x}$Sn$_x$C compounds \cite{Dias2015,Dias2018}. Local structural distortions around Mn in Mn$_3$GaC and Mn$_3$SnC leads to long and short Mn--Mn bonds. X-ray absorption fine structure (XAFS) studies have shown that the separation between the long and short Mn--Mn distances is much wider in Mn$_3$GaC \cite{Dias2017} compared to that in Mn$_3$SnC \cite{Dias201548}. Such a separation in Mn--Mn distances has an important implication on the magnetic ground state in the two antiperovskite compounds and can be related to the strain imposed on the Mn$_6$C octahedra by different types of A site atoms. \emph{Thus, though the crystal structure is cubic in both the phases, the local symmetry around Mn atoms is different and also leads to different thermal variation of Mn--Mn bond distances as compared to lattice constant.}
Stoichiometry of the B site atom also critically affects the magnetic ground state of Mn$_3$AC and Mn$_3$AN antiperovskites. It is believed that Mn $3d$ -- C $2p$ hybridization plays an important role in strengthening the antiferromagnetic order in such compounds \cite{Tong2006}. In Mn$_3$GaC, a deficiency of C suppresses the first order antiferromagnetic transition rendering the material ferromagnetic \cite{Lewis2006,Dias2014}. A systematic alteration of B site composition allows for a control over the magnetic interactions as well as hysteresis at the first order transition \cite{Cakir2013,Born2018}. \emph{Thermal expansion as well as magnetic ground state can be effectively tuned by Zn vacancies in Mn$_3$ZnN antiperovskites \cite{wang201285}.}

Recently, a detailed analysis magnetic properties of carbon deficient Mn$_3$SnC$_x$ ($x$ = 1, 0.95, 0.9, 0.85, 0.8) and tin deficient Mn$_3$Sn$_y$C ($y$ = 1, 0.95, 0.9, 0.85) showed that C deficiency strengthens ferromagnetic interactions at the expense of antiferromagnetic interactions while Sn deficiency tends to strengthen both, ferromagnetic and antiferromagnetic interactions \cite{Gaonkar2018}. These changes in magnetic interactions were attributed to changes in effective strain experienced by the Mn$_6$C functional units. It was hypothesized that C deficiency increases the tensile strain on the Mn$_6$C octahedra which elongate the Mn-–Mn bonds and thus strengthening ferromagnetic interactions. On the other hand, Sn deficiency seems to ease out the strain on the octahedral units resulting in more distorted Mn$_6$C octahedra compared to those in stoichiometric Mn$_3$SnC.

To verify the above scenario in carbon deficient and tin deficient Mn$_3$SnC, it is therefore necessary to gain an understanding of atomic magnetic moments on the Mn atoms as well as study the changes in Mn local structure in C and Sn deficient Mn$_3$SnC in comparison to those in the stoichiometric compound. To achieve these objectives, we have carried out temperature dependent neutron diffraction and Sn K and Mn K edge XAFS studies on Mn$_3$SnC, Mn$_3$SnC$_{0.8}$ and Mn$_3$Sn$_{0.85}$C.  It is found that carbon deficiency results in an interplay of interactions between nearest and next nearest Mn--Mn bonds while Sn deficiency presents more space for the Mn$_6$C octahedra to distort. These local structural changes affect the magnetic interactions between Mn atoms resulting in observed magnetic properties.

\section{Experimental}
Polycrystalline Mn$_3$SnC, Mn$_3$SnC$_{0.8}$ and Mn$_3$Sn$_{0.85}$C samples were prepared by solid state reaction method by thoroughly mixing the constituent elements, Mn, Sn and C in their stoichiometric ratio of \emph{3:1:1, 3:1:0.8 and 3:0.85:1 respectively}. A 15\% extra graphite powder was added to each mixture in order to make up for possible carbon losses during reaction \cite{yu200393}.  The mixture was then pressed into pellet, sealed in an evacuated quartz tube and sintered first at 1073K for 48 hours and then at 1150K for 120 hours \cite{Wang200985}. After cooling to room temperature the samples were powdered, mixed thoroughly and annealed again to obtain a homogeneous sample.  Crystal structure and phase purity of samples were confirmed via room temperature x-ray diffraction (XRD) study using a Rigaku diffractometer housed on a 18kW rotating anode x-ray generator with Cu anode. For these measurements, the generator was operated at 4kW (40 kV, 100 mA) and the scans were recorded at a speed of 1$^\circ$/minute and with a step of 0.02$^\circ$. Magnetization measurements were performed using a SQUID magnetometer in an applied field of 100 Oe. Magnetization was measured from 5K to 300K by first cooling the sample in zero applied field (zero field cooled - ZFC). Magnetization was also measured during subsequent cooling (FCC) and warming cycles (FCW) in the same applied field of 100 Oe. In order to understand the magnetic spin alignment in the three compounds, neutron diffraction patterns were recorded as a function of temperature on PD2 diffractometer ($\lambda$ = 1.2443 \AA)at Dhruva reactor, Bhabha Atomic Research Centre, Mumbai, India. \emph{The samples were first cooled down to 6K and the data were recorded during warming.} The patterns were Rietveld refined using FullProf suite \cite{Carvajal1993192}. XAFS data at the Mn K edge (6539 eV) and Sn K edge (29200 eV) were collected in transmission mode at selected temperatures between \emph{50K and 300K during warming} at the P65 beamline at PETRA III Synchrotron at DESY, Hamburg Germany. The absorbers of appropriate thickness for Mn edge were prepared by sprinkling finely powdered samples on tape and stacking such layers to optimize the edge jump. For Sn edge, pellets of a mixture of required amounts of sample powder and graphite were used. Further, the XAFS ($\chi$(k)) signal is extracted by reducing the K edge data using well established procedures in the Demeter program \cite{Ravel200512}.

\section{Result and Discussion}
\begin{figure}[htb]
\begin{center}
\includegraphics[width=\columnwidth]{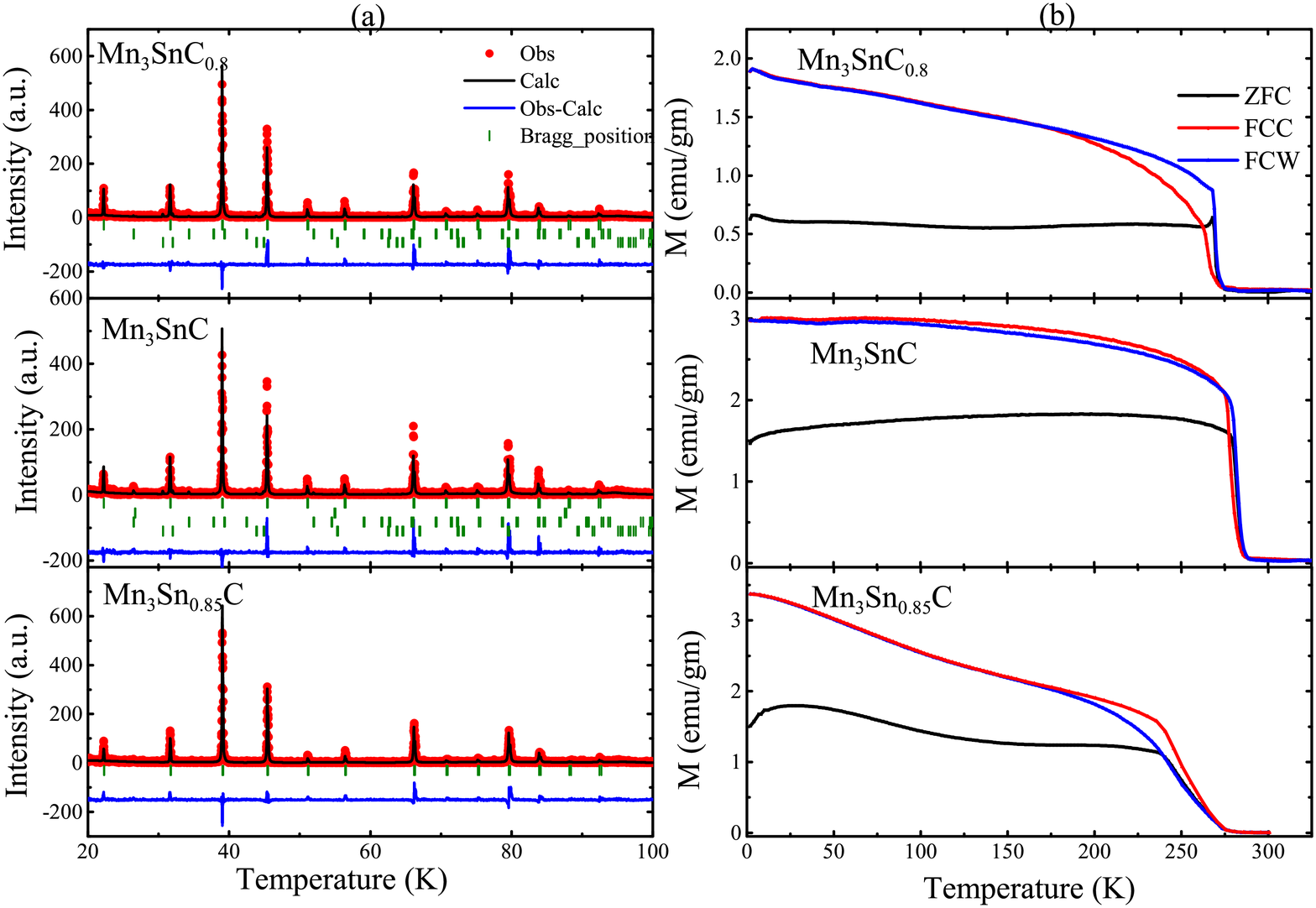}
\caption{(a) Rietveld refined x-ray diffraction patterns recorded at room temperature and (b) Magnetization as a function of temperature recorded in an applied field of 100 Oe for Mn$_3$SnC$_{0.8}$, Mn$_3$SnC and Mn$_3$Sn$_{0.85}$C.}
\label{fig:xmag}
\end{center}
\end{figure}

Figure \ref{fig:xmag}(a) presents Rietveld refined XRD plots of the three compounds, Mn$_3$SnC, Mn$_3$SnC$_{0.8}$ and Mn$_3$Sn$_{0.85}$C. It reveals formation of cubic antiperovskite phase in all three compounds along with minor impurity phases of C ($\sim$ 2\%) and MnO ($\sim$ 1\%). The value of lattice constant, estimated from the refinement, are quite similar ($a$ = 3.9890 $\pm$ 0.0005 \AA) for Mn$_3$SnC and Mn$_3$SnC$_{0.8}$ while it is slightly smaller ($a$ = 3.9840 $\pm$ 0.0005 \AA) in the case of Mn$_3$Sn$_{0.85}$C. Magnetization measured during ZFC, FCC and FCW cycles in an applied field of 100 Oe is presented in Figure \ref{fig:xmag}(b). As evidenced from hysteresis between FCC and FCW curves, all the three compounds display a first order magnetic transition from a paramagnetic state to a state with complex magnetic order. While the magnetostructural transition temperature $T_{ms}$ = 284K and the narrow hysteresis observed in Mn$_3$SnC is in agreement with that reported in literature earlier \cite{Wang200985,Dias201548}, \emph{a considerably broad hysteresis extending from $T_{ms}$ = 275K down to 170K} is seen in Mn$_3$SnC$_{0.8}$. In case of Mn$_3$Sn$_{0.85}$C, the magnetostructural transition itself is quite broad and the transition temperature, estimated from the minimum of the derivative curve is found to be $T_{ms}$ = 248K. Some hysteresis between FCC and FCW curves is also noted in case of this Sn deficient compound.

\begin{figure}[htb]
\begin{center}
\includegraphics[width=\columnwidth]{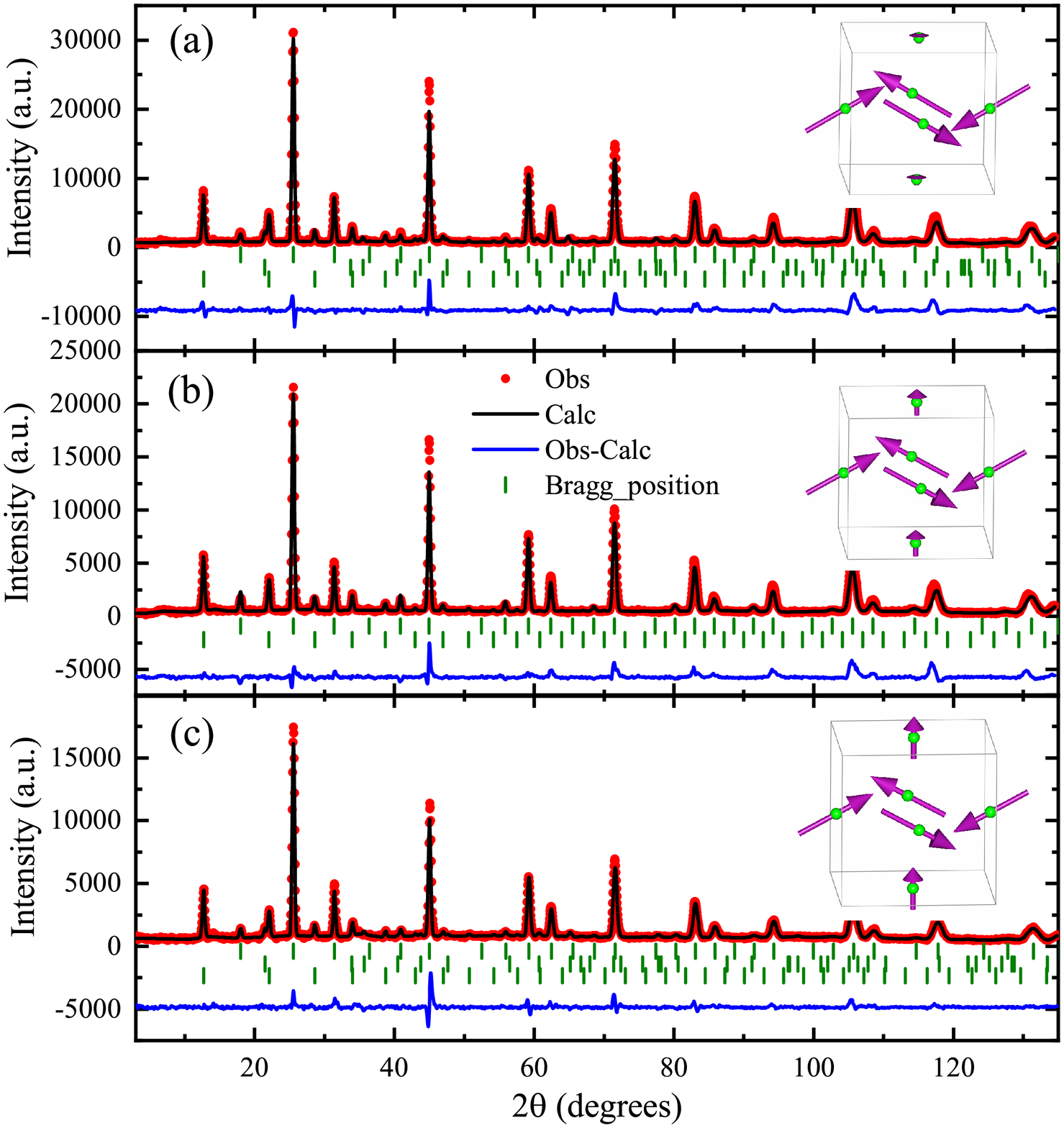}
\caption{Rietveld refined Neutron diffraction pattern of (a) Mn$_3$SnC, (b) Mn$_3$SnC$_{0.8}$ and (c) Mn$_3$Sn$_{0.85}$C recorded at 6K. The insets show the corresponding Mn spin alignment drawn within the chemical unit cell.}
\label{fig:nd}
\end{center}
\end{figure}

\begin{figure}[htb]
\begin{center}
\includegraphics[width=\columnwidth]{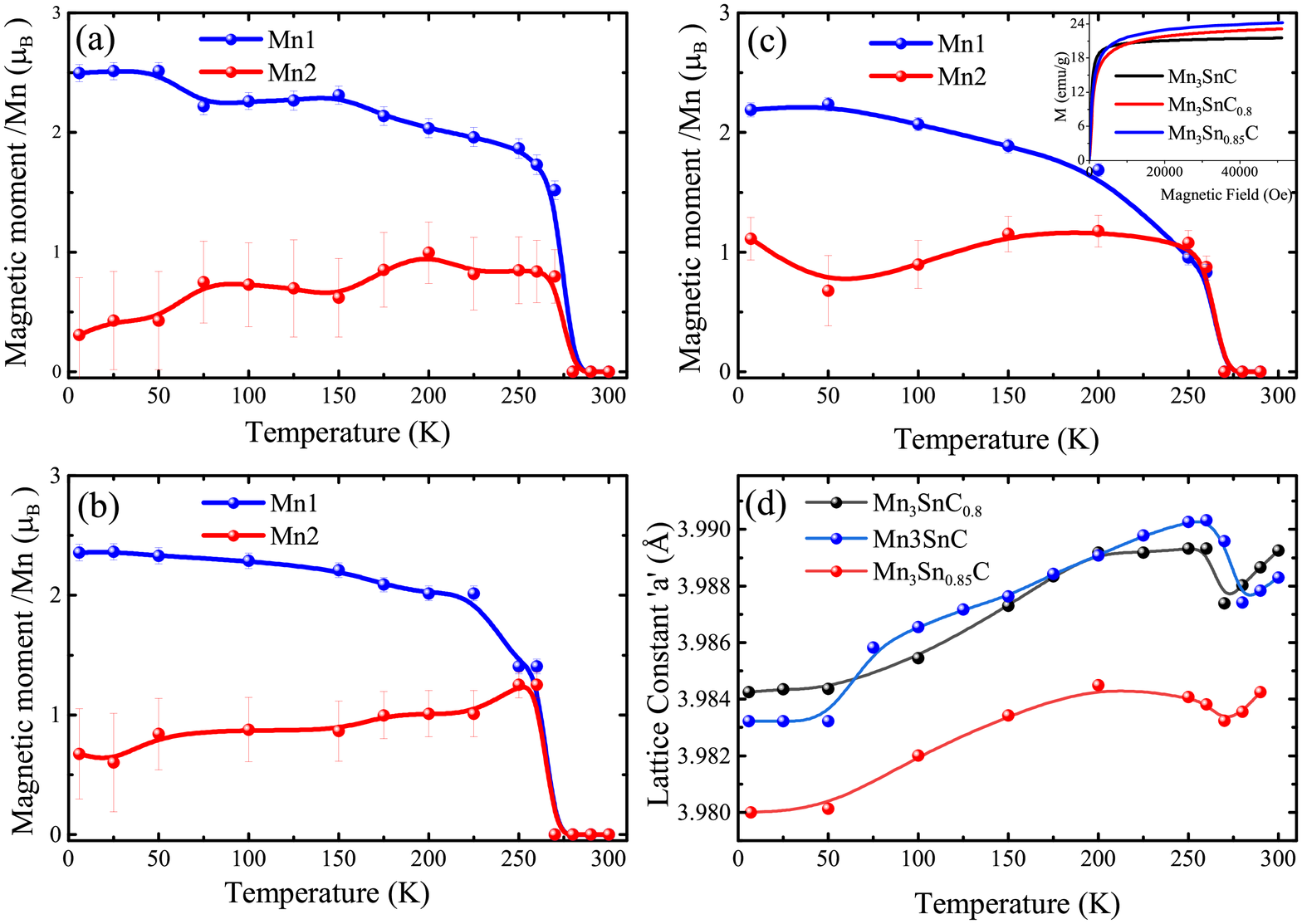}
\caption{Variation of magnetic moments Mn1 and Mn2 obtained from neutron diffraction as a function of temperature in (a) Mn$_3$SnC, (b) \emph{Mn$_3$SnC$_{0.8}$} and (c) Mn$_3$Sn$_{0.85}$C. (d) Temperature variation of lattice parameter obtained from neutron diffraction for the three compounds.}
\label{fig:moment}
\end{center}
\end{figure}

Neutron diffraction studies had highlighted the complex nature of magnetic ground state in Mn$_3$SnC. The complexity was related to the fact that there are two types of magnetically distinct Mn atoms, referred to as Mn1 and Mn2, occupying the face centered lattice positions in the Mn$3$SnC unit cell. While the moments of the Mn2 species align ferromagnetically with a moment of 0.7 $\mu_B$ along the 001 direction, those of Mn1 align antiferromagnetically with a moment of 2.4 $\mu_B$ in the x-y plane. To understand the effect of Sn and C vacancies on the alignment and magnitudes of Mn moments in the antiperovskite compound temperature dependent neutron diffraction studies were carried out. Rietveld refined neutron diffraction patterns  for Mn$_3$SnC, Mn$_3$SnC$_{0.8}$ and Mn$_3$Sn$_{0.85}$C, recorded at 6K are presented in Figure \ref{fig:nd} (a), (b) and (c) respectively. Apart from peaks arising due to long range ordering of the chemical cell, a few additional peaks, notably at 2$\theta$ = 11$^\circ$ and 21$^\circ$, corresponding to antiferromagnetic unit cell defined by propagation vector $k = {1\over{2}}, {1\over{2}}, 0$ can be seen in all three compounds. In all three cases, reasonably good fits were obtained only after considering another magnetic propagation vector $k = 0, 0, 0$ that results in a pure ferromagnetic moment on one of the three Mn atoms. Insets in each panel of figure \ref{fig:nd} depict the outline of antiperovskite cubic cell with only the magnetic (Mn) atoms at the face centered positions along with their magnetic moment vectors. The arrows and their lengths indicate the direction and the magnitude of magnetic moments of the Mn atoms.  While the directions of magnetic moments in the three compounds are similar, their magnitudes show a noticeable variation. In Mn$_3$SnC, the magnitude of the ferromagnetic moment (Mn2) is quite small, $\sim 0.4 \mu_B$/Mn at 6K while the antiferromagnetic moment (Mn1) at the same temperature is about 2.5 $\mu_B$/Mn.  Introduction of carbon vacancy results in a slight decrease in the antiferromagnetic moment to 2.3 $\mu_B$/Mn concomitant with an increase in the ferromagnetic moment to 0.7 $\mu_B$/Mn at 6K. The temperature variation of magnetic moment on Mn1 and Mn2 in Mn$_3$SnC and Mn$_3$SnC$_{0.8}$ are presented in figures \ref{fig:moment}(a) and \ref{fig:moment} (b) respectively. When compared with the thermal evolution of lattice parameters of the two compounds depicted in figure \ref{fig:moment} (d) shows that the appearance of ferromagnetic and antiferromagnetic moment in Mn$_3$SnC and Mn$_3$SnC$_{0.8}$ coincides with the sudden increase in their lattice constants indicating the transition to be a magnetostructural transition.

In case of Mn$_3$Sn$_{0.85}$C (\ref{fig:moment}(c)) the ferromagnetic moment (Mn2) shows a substantial increase to about 1$\mu_B$/Mn at 6K as compared to that in Mn$_3$SnC. The antiferromagnetic moment however decreases only marginally to about 2.2$\mu_B$/Mn from 2.5$\mu_B$/Mn in Mn$_3$SnC. Here too, the broad magnetic transition is first order in nature as the appearance of magnetic reflections in neutron diffraction pattern is accompanied by an increase in the lattice parameter.  Further, the fig. \ref{fig:moment}(d) shows that the lattice constant of Mn$_3$Sn$_{0.85}$C is smaller than those of Mn$_3$SnC and Mn$_3$SnC$_{0.8}$ and are in agreement with those reported from temperature dependent synchrotron x-ray diffraction measurements \cite{Gaonkar2018}. Thus the neutron diffraction measurements indicate that the ferromagnetic moment increases with both C and Sn vacancies and the increase being more in compound with Sn vacancies in the antiperovskite lattice. \emph{Such a trend of ferromagnetic moments obtained from neutron diffraction is in good agreement with that observed from the magnetization data \cite{Gaonkar2018}}. However, the predicted increase in antiferromagnetic moment with Sn deficiency was not seen from neutron diffraction measurements. The time dependent magnetization study as well as field variation of entropy peak indicated that while C vacancy increases the strain on the Mn$_6$C octahedra, Sn vacancy relaxes the strain and allows for a greater deformation of the Mn$_6$C octahedra \cite{Gaonkar2018}. To understand these structural changes and their implications on magnetic interactions, XAFS studies at the Mn and Sn K edges have been carried out in all the three compounds, Mn$_3$SnC, Mn$_3$SnC$_{0.8}$ and Mn$_3$Sn$_{0.85}$C.

\begin{figure}[htb]
\begin{center}
\includegraphics[width=\columnwidth]{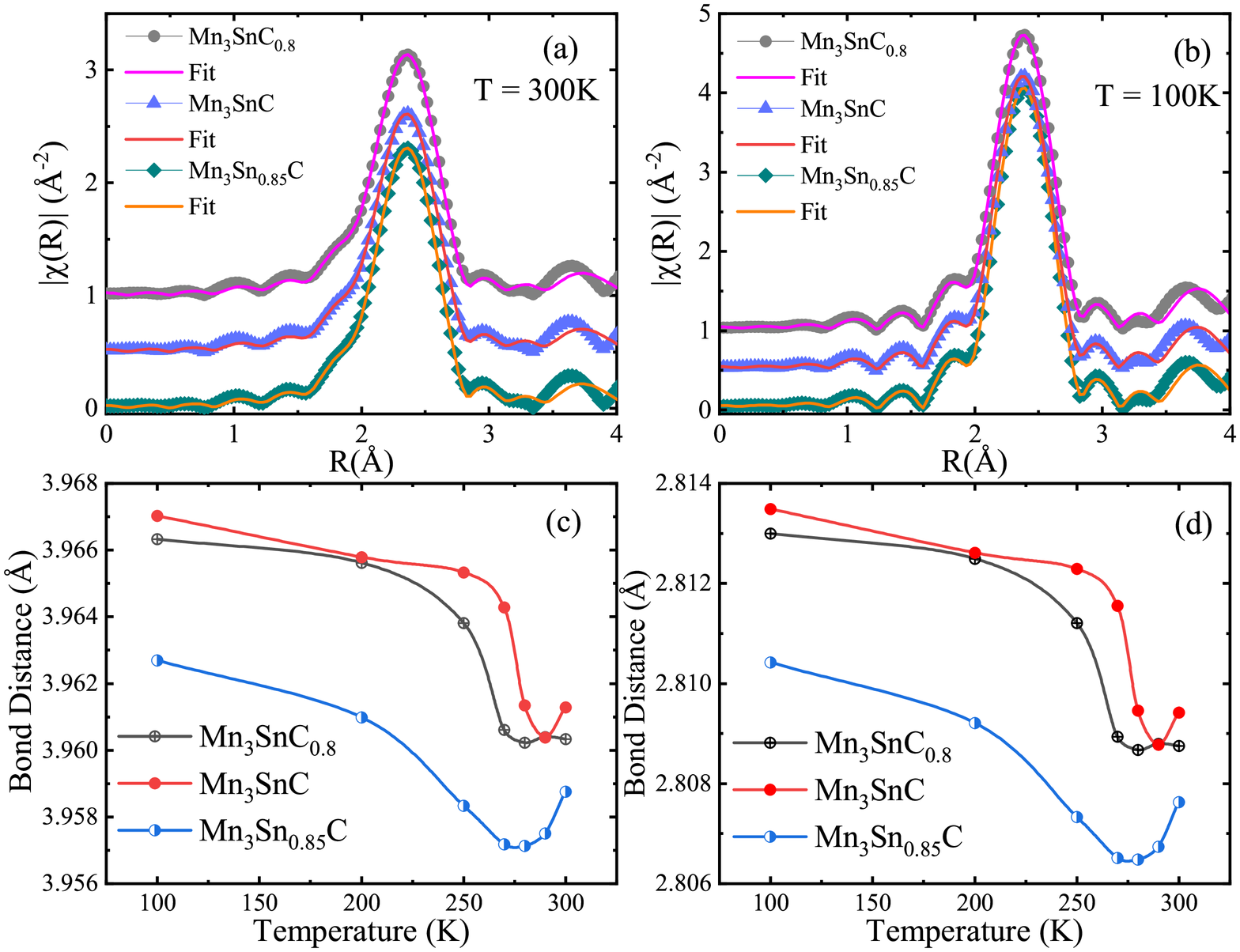}
\caption{Fourier transform of Sn K XAFS data at (a) 300K and (b) 100K in  Mn$_3$SnC$_{0.8}$, Mn$_3$SnC and Mn$_3$Sn$_{0.85}$C along with respective best fits (solid line). Temperature variation of (c) Sn--Sn and (d) Sn--Mn bond distances obtained from Sn K XAFS in Mn$_3$SnC$_{0.8}$, Mn$_3$SnC and Mn$_3$Sn$_{0.85}$C.}
\label{fig:snxafs}
\end{center}
\end{figure}

Fits to the magnitude of Fourier transform (FT) of k weighted Sn XAFS spectra recorded at 300K and 100K are shown in figure \ref{fig:snxafs}(a) and (b) respectively. For this, the XAFS data in the $k$ range of 3 to 12 \AA$^{-1}$ was transformed to $R$ space and the same was fitted in the $R$ range of 1 to 4 \AA~ around the absorbing atom (Sn) with structural correlations obtained from  the cubic structural model. In particular, the single scattering correlations between Sn--Mn, Sn--C and Sn--Sn were used in fitting. Reasonably good fits were obtained for all the three compounds at all temperatures as can be seen from the curves presented in figure \ref{fig:snxafs}(a) and (b). The bond distances obtained from the fitting also agreed well with those estimated from thermal variation of lattice constant of the three compounds, Mn$_3$SnC, Mn$_3$SnC$_{0.8}$ and Mn$_3$Sn$_{0.85}$C. Temperature variation of Sn--Mn and Sn--Sn bond lengths are presented in figures \ref{fig:snxafs}(c) and (d) respectively. These bond distances track the magnetostructural transition in the respective compounds as can be seen from the increase in Sn--Mn and Sn--Sn bond lengths at their magnetostructural transition temperature. The agreement noticed between bond distances obtained from XAFS and neutron diffraction indicate that the local symmetry around Sn atoms in all the three compounds is in consonance with the global symmetry of the lattice over the entire temperature range studied here.

\emph{Earlier study on the Mn K XAFS in Mn$_3$SnC showed that the local structure around Mn is distorted such that two of the six Mn atoms in the Mn$_6$C octahedra move away while the remaining four come closer to the central carbon atom \cite{Dias201548}}. Such a distortion results in long and short Mn--C, Mn--Mn and Mn--Sn bonds. \emph{The short Mn--Mn bond distances aid the antiferromagnetic interactions while the long Mn--Mn distances are responsible for the ferromagnetic moment on the Mn atom. Their relative weights were determined to be about 0.67 and 0.33 respectively \cite{Dias201548}.} Hence, hereinafter these short and long Mn--Mn bond distances are referred to as Mn--Mn$_{AF}$ and Mn--Mn$_F$ respectively.  However, as the Sn local structure is in accordance with cubic symmetry, all the four nearest Sn--Mn bond lengths are equal. Therefore, in the present analysis of Mn K XAFS in Mn$_3$SnC, Mn$_3$SnC$_{0.8}$ and Mn$_3$Sn$_{0.85}$C, only one unique Mn--Sn bond distance was considered and its value was held fixed to the value obtained from the analysis of Sn K XAFS in the respective compounds. It must be noted that Mn--Mn and Mn--Sn distances contribute to the same feature in the FT of Mn XAFS spectra while there is an unique peak resulting from Sn--Mn scattering in Sn XAFS. Hence Sn--Mn bond distance obtained from Sn XAFS is more accurate and also allows for a much more reliable determination of bond distance between Mn and Sn atoms.

In order to ascertain the validity of distorted local structure around Mn in Mn$_3$SnC$_{0.8}$ and Mn$_3$Sn$_{0.85}$C, Mn XAFS in these two compounds along with that recorded in Mn$_3$SnC were fitted with correlations obtained from the cubic structural model as well as a model that allowed distortions mentioned above in the Mn$_6$C octahedra (distorted octahedra model). In the cubic model, structural constraints according to cubic symmetry were imposed on Mn--C, Mn--Mn and Mn--Sn bond distances. While in the distorted octahedra model, all bond distances, except Mn--Sn bond distance, were allowed to vary freely. Apart from the unique Mn--Sn bond distance, the present distorted octahedra model consists of long and short Mn--Mn bond distances and a Mn--C bond distance. No distortions in the Mn--C bond were considered as its contribution to the total Mn XAFS is quite small.

\begin{figure}[htb]
\begin{center}
\includegraphics[width=\columnwidth]{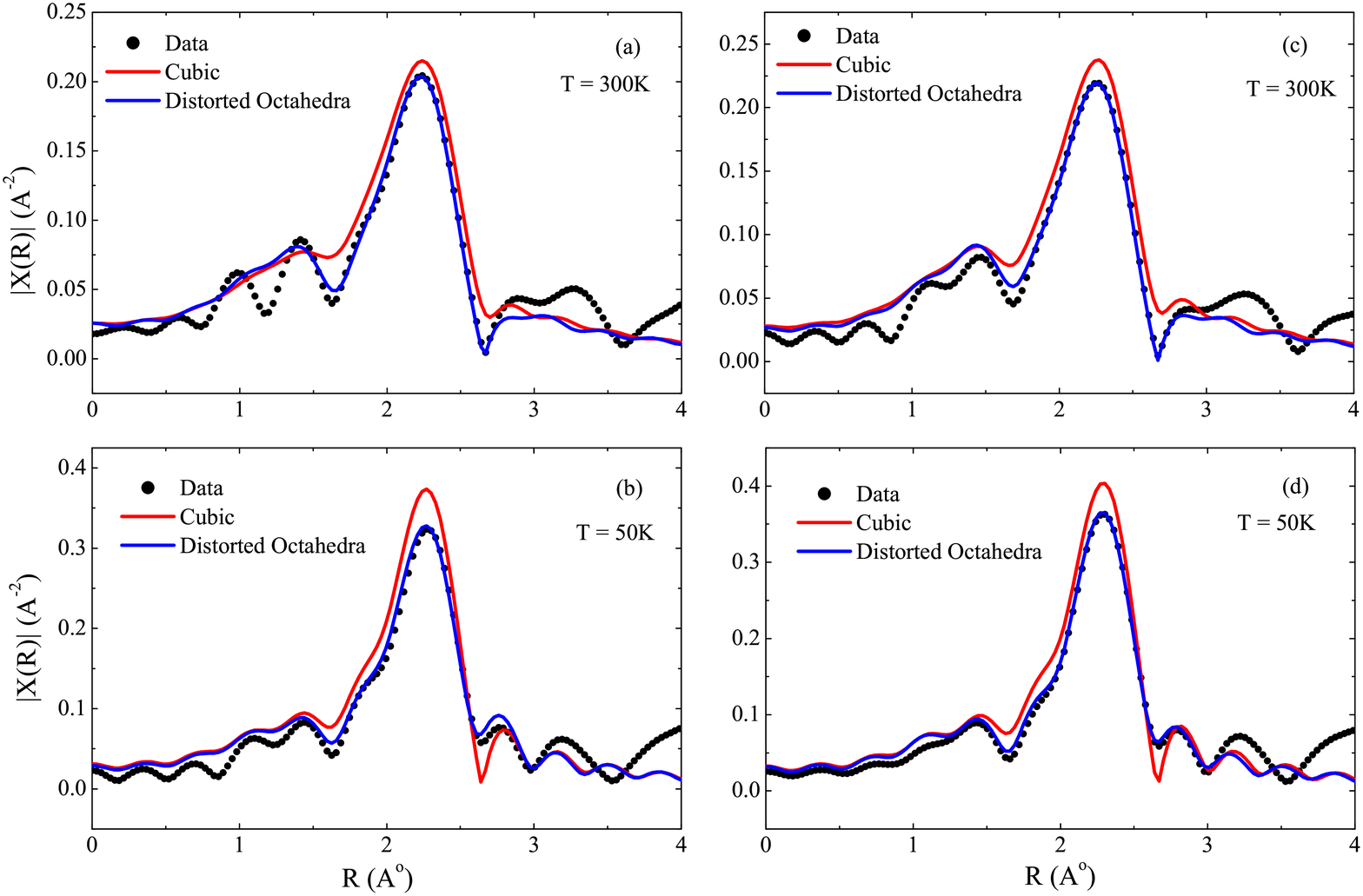}
\caption{Fourier transform of Mn XAFS data (points) along with fits (solid lines) obtained for the cubic model and the distorted octahedra model at 300K and 50K for Mn$_3$SnC$_{0.8}$ ((a) and (b) and for Mn$_3$Sn$_{0.85}$C ((c) and (d)).}
\label{fig:mnxafs}
\end{center}
\end{figure}

\begin{figure}[htb]
\begin{center}
\includegraphics[width=\columnwidth]{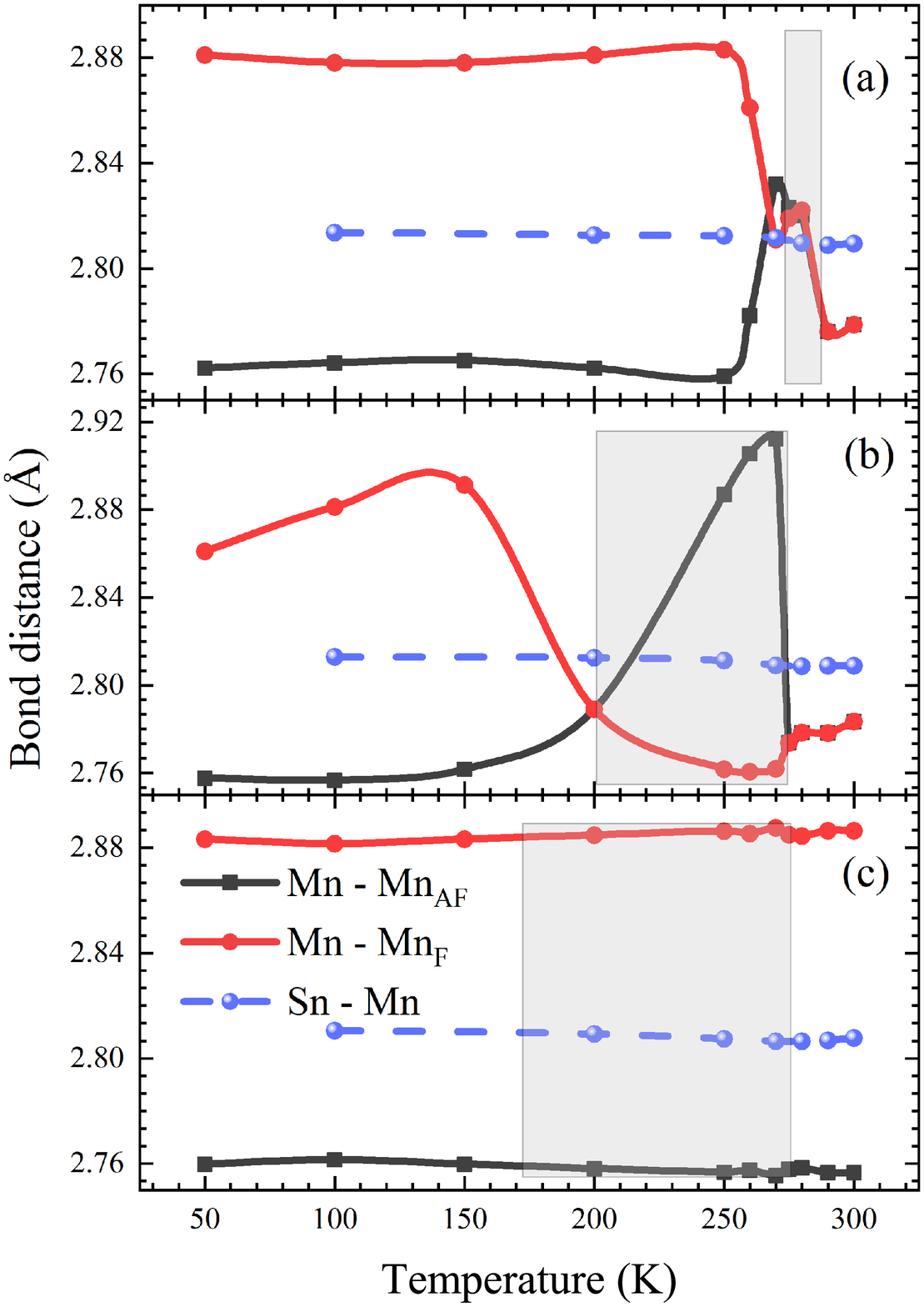}
\caption{Variation of Mn--Mn$_{AF}$ and Mn--Mn$_F$ bond distances along with Mn--Sn bond distance as a function of temperature in (a) Mn$_3$SnC, (b) Mn$_3$SnC$_{0.8}$ and (c) Mn$_3$Sn$_{0.85}$C. The shaded region indicates the transformation region for the respective compounds.}
\label{fig:mndist}
\end{center}
\end{figure}

Figure \ref{fig:mnxafs} presents fits to Mn XAFS in Mn$_3$SnC$_{0.8}$ ((a) and (b)) and Mn$_3$Sn$_{0.85}$C ((c) and (d)) at 300K and 50K respectively. Here too, XAFS data in the $k$ range of 3 to 12 \AA$^{-1}$ were Fourier transformed to $R$ space and fitted in the range of 1 to 3 \AA. Fits to Mn$_3$SnC were similar to those reported in Ref.\cite{Dias201548} and hence are not shown. It can be readily seen that the local structure around Mn in both the compounds at 50K as well as 300K is better described by the distorted octahedra model than the cubic model, implying presence of local structural distortions around Mn atoms in all the three compounds irrespective of its stoichiometry. Temperature variation of the two Mn--Mn distances along with Mn--Sn distances obtained from Sn XAFS analysis in all the three compounds are plotted in figure \ref{fig:mndist}. \emph{It must be mentioned here that Mn--Sn distances plotted here in figure \ref{fig:mndist}(a), (b) and (c) are the same as those plotted earlier in figure \ref{fig:snxafs}(d) and difference is only in the $y$-scale used.}

Variation of Mn--Mn and Mn--Sn distances in Mn$_3$SnC is shown in figure \ref{fig:mndist}(a). Although there is a general agreement with the variation and values of Mn--Mn distances reported here and those in Ref. \cite{Dias201548}, there are also some differences. These differences could be arising due to the differences in the fitting procedures employed. While only Mn XAFS was analyzed in the previous case \cite{Dias201548}, both Sn and Mn XAFS are analyzed in the present study. As mentioned earlier, such approach allows for a better estimate of Mn--Sn distance and by implication the Mn--Mn bond distances as well. The ratio of $1\over{3}$  and $2\over{3}$ determined respectively for long and short Mn--Mn bond distances \cite{Dias201548} was used in the present analysis of Mn XAFS in all three compounds. The Mn--Mn$_{AF}$ distance is assigned a weight of 0.67 and the Mn--Mn$_F$ distance is weighted by 0.33. Attempts to vary this ratio resulted in unphysical values of certain parameters.

The temperature evolution of Mn--Mn bond distances obtained from fitting Mn K XAFS spectra can be correlated with the observed magnetic properties. In Mn$_3$SnC at $T > T_{ms}$, both Mn--Mn$_{F}$ and Mn--Mn$_{AF}$ are nearly equal but shorter than Mn--Sn bond distance. At $T_{ms}$, both Mn--Mn bond distances show a sudden increase and at about 275 K, they split and move in opposite directions. At 270 K, Mn--Mn$_{AF}$ and Mn--Mn$_F$ display respective maxima and minima followed by a decrease and increase before settling down to their final values at 250 K. At all temperature below 250K Mn--Mn$_F$ is greater than Mn--Mn$_{AF}$ and have temperature invariant values of bond lengths. This variation of Mn--Mn bond distances, especially the crossover between Mn--Mn$_{F}$ and Mn--Mn$_{AF}$ bond distances in the region 290K $\ge T \ge $ 250 K augers well with the hysteresis seen between FCC and FCW magnetization curves reported in figure \ref{fig:xmag}(b).

A more or less similar variation of Mn--Mn bond distances is seen for Mn$_3$SnC$_{0.8}$ (Figure \ref{fig:mndist}(b)). The two Mn--Mn bond distances are almost equal to each other at $T > T_{ms}$ followed by a splitting that results in a much more rapid increase  Mn--Mn$_{AF}$ distance compared to the decrease in Mn--Mn$_{F}$ distance. In the entire region of transition hysteresis, as identified by hysteresis between FCC and FCW magnetization curves, Mn--Mn$_{AF}$ is greater than Mn--Mn$_F$. At the end of this transition hysteresis, the two distances cross over and below 200K, Mn--Mn$_{AF}$ is less than Mn--Mn$_F$. A comparison of the two Mn--Mn distances in Mn$_3$SnC and Mn$_3$SnC$_{0.8}$ reveals that the magnitudes of both Mn--Mn$_{AF}$ and Mn--Mn$_F$  are similar in the two samples at low temperature (T $<$ 200K). However, at T $>$ 200K, Mn--Mn$_{AF}$ distance in Mn$_3$SnC$_{0.8}$ is distinctly larger than that in Mn$_3$SnC. A larger Mn--Mn$_{AF}$ distance would imply weaker antiferromagnetic interactions and strengthening the ferromagnetic interactions. This explains the slightly higher ferromagnetic moment on Mn observed in Mn$_3$SnC$_{0.8}$ from neutron diffraction measurements.

An entirely different temperature evolution of Mn--Mn distances is witnessed for Mn$_3$Sn$_{0.85}$C. Here the two Mn--Mn bond distances are never equal, A wide separation between Mn--Mn$_{AF}$ and Mn--Mn$_F$ exists at all temperatures and Mn--Mn$_{AF}$ is always less than Mn--Mn$_F$. However, the values of the two Mn--Mn distances at 50K are similar to those in Mn$_3$SnC. The similarity of Mn--Mn bond distances in the three compounds, especially at low temperatures agrees well with the \emph{similarity of the behavior of their M(T) curves}.

A comparison of splitting between Mn--Mn$_{AF}$ and Mn--Mn$_F$ bond distances in Mn$_3$SnC and Mn$_3$SnC$_{0.8}$ reveals that the distortions in the Mn$_6$C octahedra are of two different types. In the high temperature phase ($T > T_{ms}$), both Mn--Mn$_{AF}$ and Mn--Mn$_F$ have similar values but less than the Mn--Sn bond distance. Below $T_{ms}$, the two distances show a wide separation. Though at the lowest temperature examined here, 50K, Mn--Mn$_F$ is greater than Mn--Mn$_{AF}$, in the region of hysteresis Mn--Mn$_F$ is shorter than Mn--Mn$_{AF}$ in both Mn$_3$SnC and Mn$_3$SnC$_{0.8}$. It may be recalled that in this region of transformation, isothermal magnetization loops display a crossover followed by inverted behavior at low applied magnetic field \cite{Gaonkar2018}.  While the separation between the two Mn--Mn bond distances remains nearly constant in Mn$_3$SnC below the region of hysteresis, in Mn$_3$SnC$_{0.8}$ the separation first increases and then it shows a tendency to decrease. In Mn$_3$Sn$_{0.85}$C the difference between the two Mn--Mn bond distances remains nearly constant over the entire temperature range studied here. At 50K, the separation between Mn--Mn bond distances is the least in Mn$_3$SnC$_{0.8}$ among the three samples studied here. \emph{Thus the temperature variation of Mn--Mn bond distances is much more interesting in the C deficient compound than the Sn deficient compound}.

Carbon deficiency tends to disturb the equilibrium of octahedral distortions. \emph{Due to vacancy at the center of Mn$_6$C octahedra, the nearest neighbor and next nearest neighbor Mn--Mn interactions compete with each other causing a dynamical situation that does not allow the octahedral distortions to settle resulting in the observed thermal evolution of Mn--Mn distances and a large region of hysteresis in magnetic properties. This is also revealed by the inverted M(H) loops and time dependent magnetization measurements reported in \cite{Gaonkar2018} as well as the large hysteresis in M(T).  On the other hand, the cubic cage formed by Sn atoms restrict the distortions of the Mn$_6$C octahedra within the unit cell preserving the long range cubic symmetry of the compound. Deficiency of Sn atoms allows for additional space for the Mn$_6$C octahedra to relax further resulting in temperature independent values of Mn--Mn bond distances over the entire temperature range. Though the M(T) measurement on Mn$_3$Sn$_{0.85}$C also shows presence of hysteresis between warming and cooling curves, the long relaxation time of the lattice as measured from the time dependent magnetization studies and the M(H) curves recorded just below the transformation temperature \cite{Gaonkar2018} supports the temperature independent variation of Mn -- Mn bond lengths. The reason for hysteresis in M(T) curves of Sn-deficient sample could be due to formation of regions with varying Sn content. Such regions transform at different temperatures leading to the observed hysteresis.} Thus the role of Sn or in general the A site atom is to provide a fixed volume cubic cage for the Mn$_6$C octahedra. The distortion of these octahedral units are limited to the free volume available depending on the size of the A site atom and its electronic interactions with neighboring atoms.

\section{Conclusion}
Thus the present study clearly brings out the role of Sn and C in modulating the strain on the Mn$_6$C octahedral units in Mn$_3$SnC type antiperovskites. While C atom acts as a pivot point about which the distortions of Mn$_6$C octahedra occur, Sn atoms form a restricting cage limiting the distortions and preserving the long range symmetry of the lattice. Absence of C at the center of the octahedra results in a competing structural interactions between nearest Mn atoms along the octahedral edges and next nearest Mn atoms along the diagonal of the Mn$_6$C octahedra resulting in increased strain and a wide variation of Mn--Mn distances as a function of temperature. This variation of Mn--Mn distances is responsible for the large thermal hysteresis in magnetization around the magnetostructural transition. On the other hand, a deficiency of Sn results in relaxation of the strain on Mn$_6$C octahedra and nearly temperature independent variation of Mn--Mn distances.

\section*{Acknowledgement}
Council for Scientific and Industrial Research (CSIR), New Delhi is gratefully acknowledged for financial assistance under 03(1343)/16/EMR-II. Portions of this research were carried out at the light source PETRA III at DESY, a member of the Helmholtz Association (HGF). We would like to thank Edmund Welter for assistance at beamline P65. Financial support by the Department of Science and Technology, (Government of India) provided within the framework of India@DESY collaboration is gratefully acknowledged. VNG thanks University Grants Commission for BSR Fellowship.

\bibliography{References}

\begin{thebibliography}{10}
\expandafter\ifx\csname url\endcsname\relax
  \def\url#1{\texttt{#1}}\fi
\expandafter\ifx\csname urlprefix\endcsname\relax\def\urlprefix{URL }\fi
\expandafter\ifx\csname href\endcsname\relax
  \def\href#1#2{#2} \def\path#1{#1}\fi

\bibitem{Hamada2011109}
T.~Hamada, K.~Takenaka, J. Appl. Phys. 109~(7) (2011) 07E309 (2011).

\bibitem{Takenaka2005}
K.~Takenaka, H.~Takagi, Appl. Phys. Lett. 87~(26) (2005) 261902 (2005).

\bibitem{takenaka201198}
K.~Takenaka, A.~Ozawa, T.~Shibayama, N.-h. Kaneko, T.~Oe, C.~Urano, Appl. Phys.
  Lett. 98~(2) (2011) 022103 (2011).

\bibitem{Lei201199}
L.~Ding, C.~Wang, L.~Chu, J.~Yan, Y.~Na, Q.~Huang, X.~Chen, Appl. Phys. Lett.
  99~(25) (2011) 251905 (2011).

\bibitem{Tohei200394}
T.~Tohei, H.~Wada, T.~Kanomata, J. Appl. Phys. 94~(3) (2003) 1800--1802 (2003).

\bibitem{Lewis200393}
M.-H. Yu, L.~H. Lewis, A.~R. Moodenbaugh, J. Appl. Phys. 93~(12)  10128--10130.

\bibitem{Aczel201490}
A.~A. Aczel, L.~Li, V.~O. Garlea, J.-Q. Yan, F.~Weickert, M.~Jaime, B.~Maiorov,
  R.~Movshovich, L.~Civale, V.~Keppens, D.~Mandrus, Phys. Rev. B 90 (2014)
  134403 (2014).

\bibitem{Kamishima200063}
K.~Kamishima, T.~Goto, H.~Nakagawa, N.~Miura, M.~Ohashi, N.~Mori, T.~Sasaki,
  T.~Kanomata, Phys. Rev. B 63 (2000) 024426 (2000).

\bibitem{Li200572}
Y.~B. Li, W.~F. Li, W.~J. Feng, Y.~Q. Zhang, Z.~D. Zhang, Phys. Rev. B 72
  (2005) 024411 (2005).

\bibitem{Zhang2014115}
X.~H. Zhang, Y.~Yin, Q.~Yuan, J.~C. Han, Z.~H. Zhang, J.~K. Jian, J.~G. Zhao,
  B.~Song, J. Appl. Phys. 115~(12) (2014) 123905 (2014).

\bibitem{fruchart197844}
D.~Fruchart, E.~F.~Bertaut, J. Phys. Soc. Jpn. 44~(3) (1978) 781--791 (1978).

\bibitem{kaneko198756}
T.~Kaneko, T.~Kanomata, K.~Shirakawa, J. Phys. Soc. Jpn. 56~(11) (1987)
  4047--4055 (1987).

\bibitem{takenaka201415}
K.~Takenaka, M.~Ichigo, T.~Hamada, A.~Ozawa, T.~Shibayama, T.~Inagaki,
  K.~Asano, Sci. Technol. Adv. Mater. 15~(1) (2014) 015009 (2014).

\bibitem{Wang200985}
B.~S. Wang, P.~Tong, Y.~P. Sun, X.~Luo, X.~B. Zhu, G.~Li, X.~D. Zhu, S.~B.
  Zhang, Z.~R. Yang, W.~H. Song, J.~M. Dai, Europhys. Lett. 85~(4) (2009) 47004
  (2009).

\bibitem{Dias201548}
E.~T. Dias, K.~R. Priolkar, A.~Das, G.~Aquilanti, $\rm\ddot{O}$.
  $\rm\c{C}$akir, M.~Acet, A.~K. Nigam, J. Phys. D: Appl. Phys. 48~(29) (2015)
  295001 (2015).

\bibitem{Cakir201796}
$\rm\c{C}$akir $\rm\ddot{O}$., F.~Cugini, M.~Solzi, K.~Priolkar, M.~Acet,
  M.~Farle, Phys. Rev. B 96 (2017) 014436 (2017).

\bibitem{Sun200942}
Z.~H. Sun, X.~Y. Song, F.~X. Yin, L.~X. Sun, X.~K. Yuan, X.~M. Liu, J. Phys. D:
  Appl. Phys. 42~(12) (2009) 122004 (2009).

\bibitem{Wang2012}
B.~S. Wang, W.~J. Lu, S.~Lin, J.~C. Lin, P.~Tong, B.~C. Zhao, W.~H. Song, Y.~P.
  Sun, J. Magn. Magn. Mater. 324~(5) (2012) 773--781 (2012).

\bibitem{Dias2015}
{E. T. Dias and K. R. Priolkar and $\rm\ddot{O}$. $\rm\c{C}$akir. and M. Acet,
  and A. K. Nigam}, J. Appl. Phys. 117~(12) (2015) 123901 (2015).

\bibitem{Dias2018}
E.~T. Dias, A.~Das, A.~Hoser, S.~Emura, A.~K. Nigam, K.~R. Priolkar, J. Appl.
  Phys. 124~(15) (2018) 153902 (2018).

\bibitem{Dias2017}
E.~T. Dias, K.~R. Priolkar, R.~Ranjan, A.~K. Nigam, S.~Emura, J. Appl. Phys.
  122~(10) (2017) 103906 (2017).

\bibitem{Tong2006}
P.~Tong, Y.~Sun, B.~Zhao, X.~Zhu, W.~Song, Solid State Commun. 138 (2006)
  64--67 (2006).

\bibitem{Lewis2006}
L.~H. Lewis, D.~Yoder, A.~R. Moodenbaugh, D.~A. Fischer, M.-H. Yu, J. Phys.:
  Condens. Matter 18~(5) (2006) 1677--1686 (2006).

\bibitem{Dias2014}
E.~Dias, K.~Priolkar, A.~Nigam, J. Magn. Magn. Mater. 363 (2014) 140--144
  (2014).

\bibitem{Cakir2013}
{\"{O}}.~{\c{C}}akir, M.~Acet, J. Magn. Magn. Mater. 344 (2013) 207--210
  (2013).

\bibitem{Born2018}
N.-O. Born, L.~Caron, F.~Seeler, C.~Felser, J. Alloys Compd. 749 (2018)
  926--930 (2018).

\bibitem{wang201285}
C.~Wang, L.~Chu, Q.~Yao, Y.~Sun, M.~Wu, L.~Ding, J.~Yan, Y.~Na, W.~Tang, G.~Li,
  et~al., Phys. Rev. B 85~(22) (2012) 220103 (2012).

\bibitem{Gaonkar2018}
V.~Gaonkar, E.~Dias, A.~B. Dey, R.~P. Giri, A.~Nigam, K.~Priolkar, J. Magn.
  Magn. Mater. 471 (2019) 215 -- 219 (2019).

\bibitem{yu200393}
M.-H. Yu, L.~Lewis, A.~Moodenbaugh, J. Appl. Phys. 93~(12) (2003) 10128--10130
  (2003).

\bibitem{Carvajal1993192}
J.~R. Carvajal, Phys. B 192~(1-2) (1993) 55--69 (1993).

\bibitem{Ravel200512}
B.~Ravel, M.~Newville, J. Synchrotron Radiat. 12~(4) (2005) 537--541 (2005).

\end{thebibliography}
\bibliographystyle{elsarticle-num}

\end{document}